# Fabrication and characterisation of CdSe photonic structures from self-assembled templates

G.Vijaya Prakash*, Rahul Singh, Ashwani Kumar, Rashmish K. Mishra

*Department of Physics, Indian Institute of Technology Delhi, New Delhi 110 016, India*

**Abstract**
We demonstrate here a method to fabricate CdSe photonic crystal from a very cheap fabrication route of templated self-assembly. The hexagonal close-packed photonic crystals are formed by the electrochemical growth of CdSe through the interstitial spaces between polymer nano/micro sphere templates. The confocal voids containing photonic crystals can be made either interconnected or well separated, with high uniformity. Structural and optical characterization confirms the good quality of electrochemically grown CdSe. These cheaply fabricated 2D photonic crystals provide a wide range of opportunities for optoelectronic devices.

## 1. Introduction

The analogy between the electromagnetic wave propagation in periodically arranged dielectric materials, known as photonic band-gap materials or photonic crystals(PCs), and electron waves in atomic crystals has motivated many scientists to explore new possibilities in the field of photonics [1,2]. With the advent of sophisticated microelectronic technologies, such as lithography and etching techniques, many exciting photonic crystals have been fabricated and realised as optoelectronic functional materials, covering a wide region of interest in the electromagnetic spectrum [3-5]. However, commercial viability is yet to be realised, because this *top-down* technology is very expensive and laborious and is generally not affordable. A very cheap approach through self-assembly is a feasible alternative and has gained great importance in the last few years, covering a wide variety of photonic applications [6-8]. This simple *bottom-up* technology is based on the natural self-assembly of templates and subsequently space filling the voids either by precipitation via chemical routes or by the electrochemical reduction of materials. The latter, electrochemical deposition, has several advantages; for example, the growth is from the bottom and hence ensures the complete space filling with high density materials, without any need of further processing.

*Corresponding author:
 Email:prakash@physics.iitd.ac.in, Ph: 91(11)2659 1326

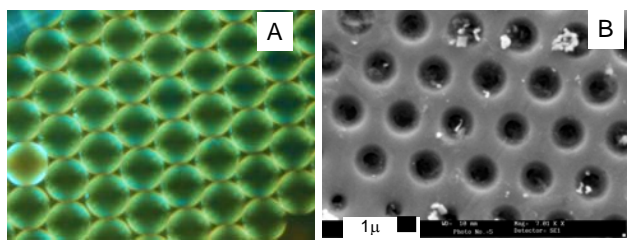

Fig.1 ( colour online). (A) High-resolution optical microscope image of Hexagonal close packed polymer microspheres ( 3μm diameter ) and (B) SEM image of CdSe photonic crystal, formed by CdSe grown upto thickness of 420nm on the templates of Fig1A and subsequently the spheres are removed .

Recently Bartlett *et al.,* reported a novel methodology to synthesize micro and nano porous meta-materials using self-assembly of polymer spheres followed by electrochemical deposition[9]. Using this low-cost fabrication technique, one can now fabricate a range of sizes, from 50 nm to 50 microns, in both ordered and disordered fashions. These micro/nanostructures have potential applications such as tunable plasmonic band gaps[10], novel types of liquid crystal displays[11], as well as for nanolaser cavities[12]. These macroporous materials are already available in the market as surface enhanced Raman scattering (SERS) devices [13]. Despite the fact that the methodology offers highly-ordered structures with very large single crystalline domains, it has so far been restricted only to metals. Attempts were reported in the very recent past on the fabrication of semiconductor Photonic pillars using interference lithography[14] and electron beam lithography[15], followed by electrochemical deposition. However, the fabrication of self-assembled semiconductor PCs is still scarce, particularly for semiconductors with a large refractive index contrast and adequate interconnectivity between the pores.

In this letter, we report a novel route for II-VI semiconductor (CdSe) photonic crystal fabrication using templated self-assembly. These photonic structures are characterised by means of optical studies and we have also attempted to obtain analytical solutions for electromagnetic wave propagation in these structures, as a suggestive exploration of photonic behaviour. This simple and effective method of fabrication paves a further way to engineer many structures of nano and micro dimensions and, most importantly, to integrate devices onto chips for further optoelectronic applications.

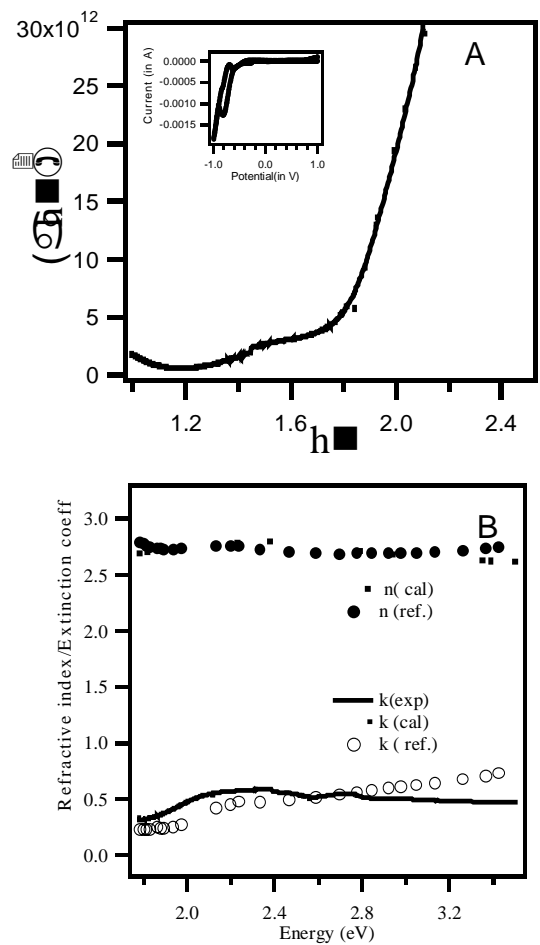

Fig.2 (A) Tauc plot of CdSe thin film ( dotted line is a theoretical fit). (B) refractive index (*n*) and extinction coefficients (*k*) of CdSe thin film (cal= calculated data (see text); exp=experimental data and ref=data taken from ref.28); Inset of Fig 2A is the typical cyclic voltagram of CdSe electrolyte with reference to reference electrode potential. Below -0.8V and above -0.4V regions are for bulk Cd and Se depositions respectively.

## 2. Experiment

Preparation of semiconductor photonic crystals is a two-step process; templating of polymer spheres and electrochemical deposition. Templates are prepared on ITO coated glass using a sedimentation method [9]. Briefly, the colloidal solutions (aprox.3% in water) of 3 μm polymer spheres (Duke Scientific Co.), are dispersed on a pre-defined area (typically 1cm$^2$ area) of ITO and subsequently the solution is allowed to evaporate slowly under controlled humidity and temperature. The assembly of spheres could be controlled either giving one layer or several layers (so called artificial opals), by simply adjusting polymer sphere concentration. These templates are very robust and adhere well to the substrate. High-resolution optical microscope images clearly show that the templates were spontaneously organised into a closely packed hexagonal structure (Fig1A). Though these structures are visible as hexagonal close packed structure (hcp) from the top, several layers of polymer sphere assembly results into a face-centered cubic geometry (fcc).

Electrochemical deposition has been carried out by a three-electrode method, using a potentiostat/galvanostat (Princeton Applied Research, Model 263A). An electrolytic mixture of $CdSO_4$ and $SeO_2$ along with $H_2SO_4$ in deionised water is used, according to ref.16. To ensure the uniformity and stoichiometry of deposited CdSe, we attempted both potentiostatic as well as galvanostatic methods to ensure finer deposition. After several experimentations, we found -0.8V constant potential, under controlled bath temperatures, is suitable for depositing good quality nearly-stoichiometric CdSe. UV-Visible absorption spectra (Perkin-Elmer Lambda 900), Glancing Angle X-ray Diffraction(Phillips, X'part pro), X-ray Fluorescence, Scanning electron microscopy (SEM, LEO 435VP), High-resolution optical microscopy (LYCOS) and thickness profilometry (AMBIOS, XP series) are employed to ensure the chemical composition, nature of the structure, surface morphology and the smoothness of the materials.

## 3. Results and Discussion

X-ray diffraction and X ray fluorescence confirm the characteristic polycrystalline hexagonal (wurtzite) phase and nearly-stoicheometry of CdSe. Our present and previous observations suggest that the reduction of materials depends strongly on the conducting nature of the substrate, as well as the surface quality. Though the growth of films of upto 1μm thickness is possible on ITO, for thicker films one has to resort to pre-treated (surface modification by higher-order silanes or surfactants) metal (such as gold and silver) coated substrates, which enhances the quality of the deposition. Moreover, the deposition and composition of CdSe is severely dependent on the experimental conditions such as potential, pH and temperature of the bath. For the deposition of CdSe on to the template, we left the templated substrate in the electrochemical bath for several minutes, to ensure the total wetting of the template. The CdSe growth takes place in the interstitial spacing between the polymer spheres. Eventually after the deposition, polymer spheres are removed by dissolving in an organic solvent, leaving the CdSe photonic crystal. One of the potential advantages of the technique is that, by simply altering thickness, one could get 2D photonic void array or by growing several layers of sphere height a 3D photonic crystal. Unlike other templating techniques, this method facilitates exact 'cast' of the spheres without any strain, which is usually a problem in other templating methods. The scanning electron microscope image (Fig.1B) shows the resultant 2D hexagonal PC, where the pore diameter is 2.1μm, corresponding to 420nm CdSe film growth. We are able to produce several pore sizes, from semi opening (half the diameter growth where the pores are interconnected) to almost closed hollow sphere structures. However, about 10% of the bottom area of the individual spherical casts is completely flat where the electrochemical growth appears to be screened under the templated spheres. This is due to the physical contact of the spheres with the substrate as well as due to the fact that the electrochemical solution is inhibited from creeping underneath the spheres. Such flatness could be minimised by appropriately controlling the deposition parameters.

It is obvious that the quality of photonic devices is potentially dependent on the optical and structural properties of the material prepared under various experimental conditions. Moreover, selenides are generally prepared from solid state reactions and, therefore, we studied our electrochemically deposited CdSe thin film prior to the fabrication of photonic crystals. The optical band gap (Fig.2A) and the edge of the Urbach energy of the CdSe thin-film are estimated from the Tauc and Urbach plots respectively, using UV-visible absorption spectra [17]. Urbach energies, the absorption tails extended in the vicinity of band gap of bulk CdSe, generally appear due to impurities and defect related absorptions. Both the band gap and Urbach energies obtained from these fits, 1.75eV and 0.45eV respectively, are very close to the values of CdSe thin-films reported in the literature[18,19]. We also obtained optical constants, refractive index($n$) and extinction coefficients ($k$), directly from the absorption data using the improvised approach proposed by Forhui and Bloomer[20,21]. It is clear from the plots (Fig.2B) that the values obtained by this method are reasonably close to the standard CdSe values.

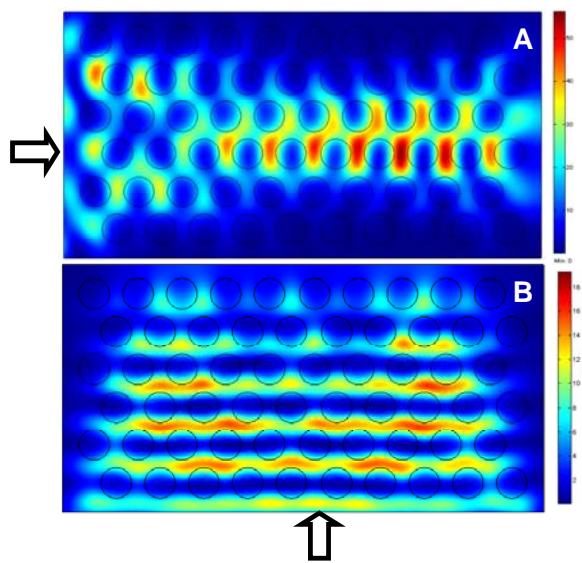

Fig.3. (colour online). Electric field intensity distribution mappings of (2D) CdSe photonic crystal for two cases, where the field propagation is (A) from left to right (B) and from bottom to top, at λ=12μm for TE mode. Arrows indicate the source boundary.

In order to emphasize the importance of photonic applications, we obtained analytical solutions for electromagnetic light propagation in these structures using finite element analysis [22]. For computational simplicity, we confined our model to the 2D domain, where air-filled ($n = 1$) circles are embedded in CdSe ($n = 2.45$). The geometry (33x17μm) contains six layers of hexagonally packed pores of similar dimensions to our CdSe PC. The solutions are obtained for two cases: (1) when the electric field source as plane wave is fed through the entire length of left boundary and another (2) through the bottom. The remaining boundaries of the geometry are set as low-reflecting. We solved the entire region of the domain using TE polarised plane wave and computed the electric field along the *z*-direction with the minimum mesh size of 300nm. Figure 3A and 3B represents the electric field intensity, mapped over the domain for electromagnetic waves of wavelength 12μm, for left–to-right and bottom-to-top electric field propagations, respectively. The transmission spectra thus obtained over the infrared region is plotted as a function of size parameter, $a/\lambda$ (where $a$ is the air pore radius) in figure 4. Transmission spectra demonstrate quite complex photonic modes, which are essentially not observable in bulk CdSe. However, our numerical solutions are too limited to totally describe photonic nature, as high-index contrast structures require complete 3D calculations of the photonic bands. A more rigorous numerical approach and experimental transmission/reflection studies, including both angular tuning and wave guiding, are being taken up as further study.

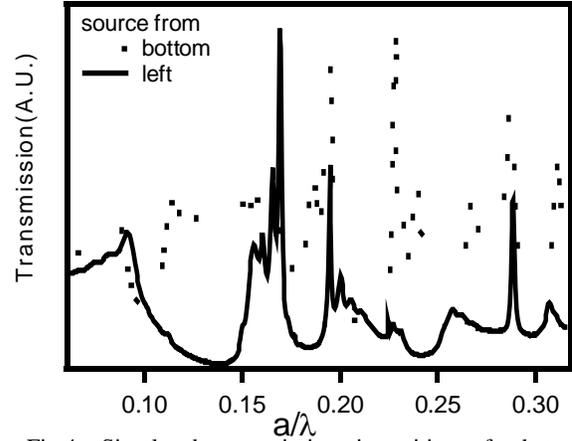

Fig.4. Simulated transmission intensities of photonic crystals as a function of size parameter ($a/\lambda$), where $a$ is pore radius. Dotted and solid lines indicate the TE light mode propagation from left and bottom boundaries respectively ( see Fig.3).

As such these 2D-PCs open an opportunity for a wide window of interesting studies: for example, waveguide applications as distributed photonic scatters (DPS)[23,24], and as active photonic crystals by filling these PCs with optically sensitive materials such as liquid crystals[11,25], quantum dots[26] and j-aggregates[27] for light-matter interactions. Also this simplest method could be used to fabricate 3D structures of metal alloys, metallic superlattices or of metal-semiconductor multi-layers.

## 4. Conclusion

In conclusion, we have successfully demonstrated a very simple and low cost method of fabricating CdSe semiconductor photonic crystals using self-assembly templates. The quality of wurtzite CdSe is confirmed by structural and optical studies. Our results therefore clearly demonstrate how low-cost fabrication techniques can be used to produce good quality CdSe 2D PCs, with potential applications inoptoelectronic devices.

## 5. Acknowledgment